\documentclass[twocolumn,preprintnumbers,amsmath,amssymb,prl]{revtex4}
\usepackage{graphicx}
\usepackage{pictexwd}
\usepackage{dcolumn}
\usepackage{amsmath}
\usepackage[usenames,dvipsnames]{xcolor}
\newcommand{\synopsis}[1]{}
\newcommand{\noopsort}[1]{}

\newcommand{\oldversion}[1]{}

\begin{document}
\title[Short Title]{Curvature Elasticity of the Electric Double Layer}

\author{Guilherme Volpe Bossa$^{1}$, Bjorn K. Berntson$^2$ and Sylvio May$^1$}
\affiliation{$^{1}$ Department of Physics, North Dakota State University, 
Fargo, ND 58108-6050\\
$^{2}$  Department of Mathematics, North Dakota State University, 
Fargo, ND 58108-6050}

\date{\today}

\begin{abstract}
Mean-field electrostatics is used to calculate the bending moduli of an electric double layer for fixed surface charge density of a macroion in a symmetric 1:1 electrolyte. The resulting expressions for bending stiffness, Gaussian modulus, and spontaneous curvature refer to a general underlying equation of state of the electrolyte, subject to a local density approximation and the absence of dipole and higher-order fields. We present results for selected applications: the lattice-gas Poisson-Fermi model with and without asymmetric ion sizes, and the Poisson-Carnahan-Starling model.
\end{abstract}

\maketitle
Electrolytes neutralize the charge carried by embedded macroions through the formation of a diffuse ion cloud, enriched in counterions and depleted in coions. This composite structure -- referred to as the electric double layer (EDL) -- is ubiquitous in cellular biology and impacts a multitude of technological applications such as supercapacitors for energy storage \cite{simon14}, capacitive deionization \cite{welgemoed05}, transport in nanofluidics \cite{sparreboom09}, drug delivery and medical imaging \cite{frohlich12}. 
The classical mean-field model of the EDL is known as Poisson-Boltzmann (PB) theory; refined models account for ion size and structure, solvent properties, ion correlations, and specific ion-ion interactions \cite{vlachy99,benyaakov09,bohinc17}. Most of these focus on the planar geometry. However, electrified interfaces are often curved or undergo bending fluctuations. Among the numerous examples are nanoporous electrodes for supercapacitor applications \cite{wang12},  charged microemulsions \cite{peira08}, biomembrane remodeling by proteins and peptides \cite{yang07,graham10}, complex formation of curved macroions such as cationic membranes and DNA \cite{hochrein07}, and fluctuation-induced topological phase transitions of model membranes \cite{golubovic94,morse94}.

The dependence of EDL structure and energy on curvature can be described in the limit of small bending by a set of curvature elastic constants. In two seminal papers, Lekkerkerker \cite{lekkerkerker89a,lekkerkerker90} has employed two different approaches (the first is a charging method and the second the determination of the lateral pressure profile) to calculate the contribution of the EDL to the curvature elastic constants based on the classical PB model. Subsequent studies have generalized these results to account -- still within the PB framework -- for curvature-dependent surface charges, modifications in the dielectric constant, and confined geometries \cite{winterhalter92,may96,fogden97}. Yet, attempts to compute the curvature elastic constants for models that go beyond the PB level are largely missing.

In the present work we apply the charging method to a class of models that, unlike the classical PB model, include a nonideal mixing contribution of the mobile ions. Our mean-field approach is used to obtain the curvature elastic constants directly from the underlying equation of state of the electrolyte. We present a general formalism and discuss three examples: the lattice-gas PB approach (which, following a suggestion by Kornyshev \cite{kornyshev07}, we refer to as the Poisson-Fermi model) with and without equal sizes of the mobile cations and anions, and the Poisson-Carnahan-Starling model that employs the Carnahan-Starling equation of state for size-equal ions. 

Consider a single macroion of surface charge density $\sigma$ immersed in a symmetric 1:1 electrolyte of bulk ion concentration $2 \phi_0/\nu$, where $\nu$ is the effective volume per salt ion and $\phi_0$ the bulk volume fraction of each individual ion type. We describe the EDL that builds up in the electrolyte outside the macroion by a mean-field self-consistency relation 
\begin{equation} \label{hu35}
l^2 \nabla^2 \Psi=f(\Psi) f'(\Psi)
\end{equation}
for the dimensionless electrostatic potential $\Psi=e \Phi/k_BT$, where $\Phi$ denotes the electrostatic potential, $e$ the elementary charge, $k_B$ Boltzmann's constant, $T$ the absolute temperature, and $l$ is a characteristic length. The function $f(\Psi)$ (with its derivative $f'(\Psi)=df/d\Psi$) depends on the underlying equation of state of the electrolyte and can be calculated from the right-hand side of Eq.~\ref{hu35} through
\begin{equation} \label{ou41}
f(\Psi)=\pm \sqrt{2 \int \limits_0^\Psi d\bar{\Psi} f(\bar{\Psi}) f'(\bar{\Psi})}.
\end{equation}
We do not consider cases where $f$ depends explicitly on any spatial derivatives of $\Psi$ -- this effectively excludes models beyond the local density approximation and confines us to ions carrying simple point charges. Examples that go beyond Eq.~\ref{hu35} include higher-order Poisson-Boltzmann equations \cite{bazant11,blossey17} and the dipolar Poisson-Boltzmann approach \cite{abrashkin07}. Nevertheless, Eq.~\ref{hu35} embodies a range of frequently used models for electrolytes with varying ion sizes, shapes, and non-electrostatic ion-ion interactions.

The free energy of the EDL can be calculated based on integrating the surface potential $\Phi$ as a function of the surface charge density $\sigma$ or, equivalently, integrating the dimensionless surface potential $\Psi$ as a function of the scaled surface charge density $s=\nu \sigma/(l e)$,
\begin{equation} \label{ft54}
F=\int \limits_A da \int \limits_0^\sigma d\bar{\sigma} \: \Phi(\bar{\sigma})=k_BT \: \frac{l}{\nu} \int \limits_A da \int \limits_0^s d\bar{s} \: \Psi(\bar{s}). 
\end{equation}
The integration $\int_A da$ runs over the macroion surface. When that surface is only weakly curved we can curvature expand the free energy and compare the resulting expression with Helfrich's free energy \cite{helfrich73} 
\begin{equation} \label{ft58}
\frac{F}{A}=\frac{F_0}{A}+\frac{\kappa}{2} (c_1+c_2)^2-\kappa c_0 (c_1+c_2)+\bar{\kappa} c_1 c_2,
\end{equation}

measured per unit area $A$, where $F_0$ is the free energy for flat geometry, and $c_1$ and $c_2$ are the two principal curvatures at a given point on the macroion surface. We calculate the bending stiffness $\kappa$, Gaussian modulus $\bar{\kappa}$, and spontaneous curvature $c_0$. Following Lekkerkerker \cite{lekkerkerker89a}, we consider Eq.~\ref{hu35} for spherical ($n=2$) and cylindrical ($n=1$) symmetry
\begin{equation} \label{xs47}
l^2 \left[\frac{d^2\Psi}{dr^2}+\frac{n}{r} \frac{d\Psi}{dr}\right]=f(\Psi) f'(\Psi)
\end{equation}
and express the radial distance $r=1/c+l x$ in terms of a dimensionless coordinate $x$ so that the macroion surface is located at $x=0$. Next, we expand $\Psi(x)=\Psi_0(x)+c l \Psi_1(x)+c^2 l^2 \Psi_2(x)$ up to quadratic order in curvature: $c_1-c=c_2=0$ for cylindrical and $c_1=c_2=c$ for spherical geometry. The result is a set of three ordinary differential equations,
\begin{eqnarray} \label{lb64}
\Psi_0''&=&f f', \hspace{1.5cm} \Psi_1''=[f f']' \: \Psi_1-n \Psi_0',\nonumber\\
\Psi_2''&=&[f f']' \:\Psi_2-n \Psi_1'+[f f']'' \: \frac{\Psi_1^2}{2} +n x \Psi_0',
\end{eqnarray}
where here and below we use the notation $f=f(\Psi_0)$, $f'=f'(\Psi_0)$, $[f f']'=f'^2+f f''$, $f''=f''(\Psi_0)$, $[f f']''=3 f' f''+f f'''$, and $f'''=f'''(\Psi_0)$. Note $\Psi_0'=d\Psi_0/dx$ and analogously for $\Psi_1'(x)$, $\Psi_2'(x)$, and higher derivatives. Because the macroion is isolated, we demand $\Psi_0(x)=\Psi_1(x)=\Psi_2(x)=0$ for $x \rightarrow \infty$. In this case, the first integration of Eqs.~\ref{lb64} can be carried out, 
\begin{eqnarray} \label{lt21}
\Psi_0'&=&-f, \hspace{2.1cm}\Psi_1'=-f' \Psi_1-n \frac{I}{f},\nonumber\\
\Psi_2'&=&-f' \Psi_2-f'' \frac{\Psi_1^2}{2}+n \left(\frac{f'}{f^2} I-1\right) \Psi_1\nonumber\\
&+&n x \frac{I}{f}+\frac{n^2}{2} \frac{I^2}{f^3}+\frac{n (1-n)}{f} \int \limits_0^{\Psi_0} d\Psi \frac{I(\Psi)}{f(\Psi)},
\end{eqnarray}
where we define $I=I(\Psi_0)=\int_0^{\Psi_0} d\Psi f(\Psi)$. For a fixed (scaled) surface charge density $s$ at the macroion surface (at $x=0$) the boundary conditions $\Psi_0'(0)+s=\Psi_1'(0)=\Psi_2'(0)=0$ must be fulfilled. When applied to $x=0$, Eqs.~\ref{lt21} yield the surface potential contributions explicitly as functions of $s$ 
\begin{eqnarray} \label{lr42}
\Psi_0(0)&=&f^{-1}(s), \hspace{1cm}
\Psi_1(0)=-n \left[\frac{I}{f f'}\right]_{\Psi_0(0)=f^{-1}(s)},\nonumber\\
\Psi_2(0)&=&\frac{n^2}{2} \left[\frac{1}{f f'} \frac{d}{d\Psi_0} \left(\frac{I^2}{f f'}\right)\right]_{\Psi_0(0)=f^{-1}(s)}\nonumber\\
&+&n (1-n) \left[\frac{1}{f f'} \int \limits_0^{\Psi_0} d\Psi \frac{I(\Psi)}{f(\Psi)} \right]_{\Psi_0(0)=f^{-1}(s)}.
\end{eqnarray}
Note that $f^{-1}(s)$ denotes the inverse function of $f$ so that $f(f^{-1}(s))=s$. The curvature contributions to the surface potential, $\Psi_1(0)$ and $\Psi_2(0)$, initially depend on $\Psi_0(0)$ -- they acquire their dependence on $s$ through the relation $\Psi_0(0)=f^{-1}(s)$. We use the surface potential contributions $\Psi_0(0)=\Psi_0(0;\bar{s})$, $\Psi_1(0)=\Psi_1(0;\bar{s},n)$, and $\Psi_2(0)=\Psi_2(0;\bar{s},n)$ in  Eq.~\ref{lr42} to determine the free energy $F$ via the charging process specified in Eq.~\ref{ft54},
\begin{equation} \label{kk87}
\frac{F}{A k_BT}=\frac{l}{\nu} \int \limits_0^s d\bar{s} \left[\Psi_0(0)+c l \Psi_1(0)+c^2 l^2 \Psi_2(0)\right].
\end{equation}
Eq.~\ref{kk87} is compared with Eq.~\ref{ft58}, both for cylindrical geometry ($n=1$), where $F/Ak_BT=F_0/Ak_BT-\kappa c_0 c+\kappa c^2/2$, and for spherical geometry ($n=2$), where $F/Ak_BT=F_0/Ak_BT-2 \kappa c_0 c+(2 \kappa +\bar{\kappa}) c^2$. This results in expressions for the bending stiffness $\kappa$, Gaussian modulus $\bar{\kappa}$, spontaneous curvature $c_0$, and free energy at flat geometry $F_0$,
\begin{eqnarray} \label{ft65}
\frac{\kappa}{k_BT}&=&\frac{l^3}{\nu} \int \limits_0^s d\bar{s} \: \left[\frac{1}{f f'} \frac{d}{d\Psi_0} \left(\frac{I^2}{f f'}\right)\right]_{\Psi_0(0)=f^{-1}(\bar{s})},\nonumber\\
\frac{\bar{\kappa}}{k_BT}&=&-2 \frac{l^3}{\nu} \int \limits_0^s d\bar{s} \: \left[\frac{1}{f f'} \int \limits_0^{\Psi_0} d\Psi \frac{I(\Psi)}{f(\Psi)} \right]_{\Psi_0(0)=f^{-1}(\bar{s})},\nonumber\\
\frac{\kappa c_0}{k_BT}&=&\frac{l^2}{\nu} \int \limits_0^s d\bar{s} \: \left[\frac{I}{f f'}\right]_{\Psi_0(0)=f^{-1}(\bar{s})},\nonumber\\
\frac{F_0}{A k_BT}&=&\frac{l}{\nu} \int \limits_0^s d\bar{s} \: f^{-1}(\bar{s}).
\end{eqnarray}
Eqs.~\ref{ft65} -- the major result of the present work -- predict the bending properties emerging from the self-consistency relation in Eq.~\ref{hu35} at any fixed surface charge density. The only input is the function $f$ (with its derivative $f'$ and integral $I$). Next, we present applications and relate $f$ to the underlying equation of state.

Classical PB theory considers point-like ions with ideal mixing properties in an electrolyte of Debye screening length $l_D=l/\sqrt{2 \phi_0}$ and Bjerrum length $l_B=\nu/(4 \pi l^2)$. The classical PB equation, $l_D^2 \nabla^2 \Psi=\sinh \Psi$, implies $f(\Psi)=2 (l/l_D) \sinh(\Psi/2)$ and thus $f'(\Psi)=(l/l_D) \cosh(\Psi/2)$, $I(\Psi)=8 (l/l_D) \sinh^2(\Psi/4)$, and $f^{-1}(s)=2 \: \mbox{arsinh} (s l_D/2 l)$. Using these in Eqs.~\ref{ft65} results in 
\begin{eqnarray} \label{li64}
\frac{\kappa}{k_BT}&=&\frac{l_D}{2 \pi l_B} \: \frac{(q-1) (q+2)}{q \: (q+1)}, \hspace{0.5cm}
\frac{\bar{\kappa}}{k_BT}=\frac{-l_D}{\pi l_B} \int \limits_{\frac{2}{1+q}}^{1} \frac{dz \ln z}{z-1}\nonumber\\
\frac{\kappa c_0}{k_BT}&=&\frac{\ln \left(\frac{1+q}{2}\right)}{\pi l_B}, \hspace{0.5cm} \frac{F_0}{A k_BT}=\frac{1-q+p \: \mbox{arsinh} \: p }{\pi l_B l_D}
\end{eqnarray}
with $q=\sqrt{1+p^2}$ and $p=s l_D/(2 l)=2 \pi l_B l_D \sigma/e$. Eqs.~\ref{li64} coincide with Lekkerkerker's results \cite{lekkerkerker89a,lekkerkerker90}.

\begin{figure}[ht!]
\begin{center}
\includegraphics[width=8.5cm]{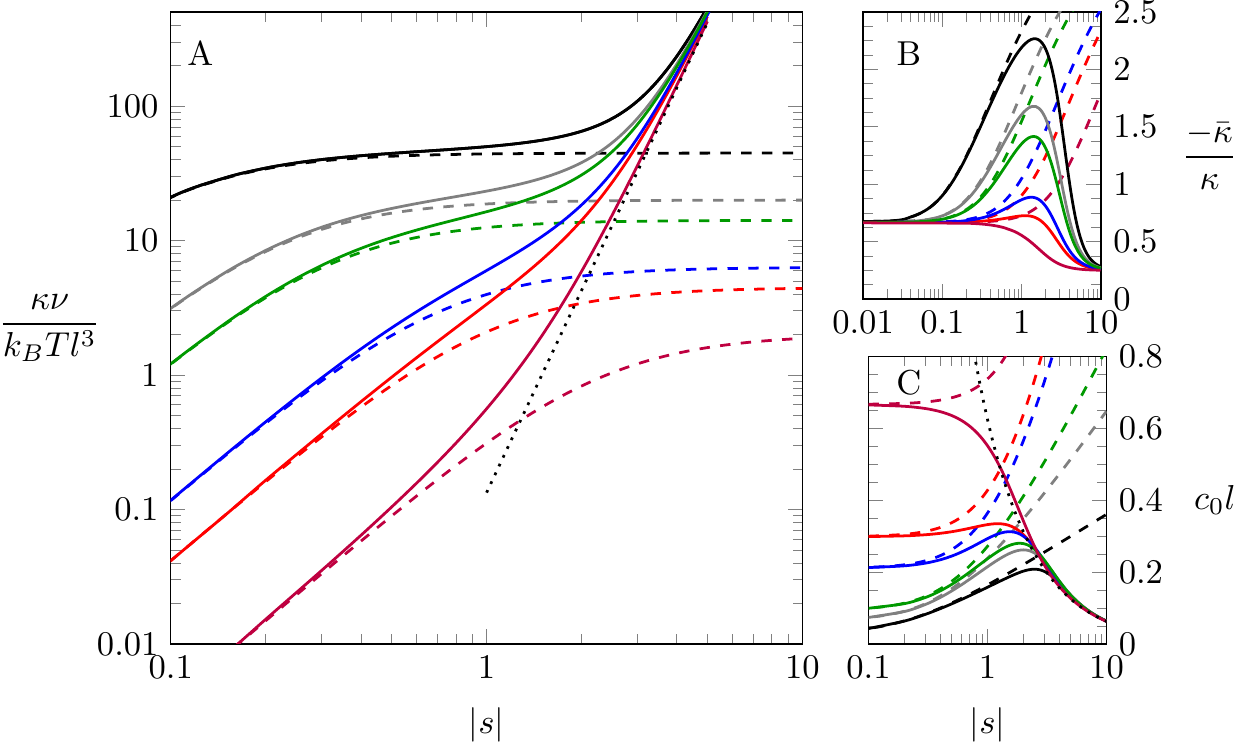}
\end{center}
\caption{\label{fig1} $\kappa/k_BT \times \nu/l^3$ (diagram A), $-\bar{\kappa}/\kappa$ (B), and $c_0 l$ (C), for $\phi_0=0.5, 0.1, 0.05, 0.01, 0.005, 0.001$ (purple, red, blue, green, grey, black) according to the Poisson-Fermi model (Eqs.~\ref{fg54} and \ref{hh70}, solid lines) and the classical PB limit (Eqs.~\ref{li64}, dashed lines). The black dotted lines mark the large-$s$ limit.
}
\end{figure}

An approximate method to account for the non-vanishing volume $\nu$ of the mobile salt ions is based on the mixing properties of a lattice-gas, which leads to the Poisson-Fermi equation \cite{borukhov97,kornyshev07},
\begin{equation} \label{fg54}
l^2 \nabla^2 \Psi=\frac{2 \phi_0 \sinh \Psi}{1+2 \phi_0 (\cosh \Psi -1)},
\end{equation}
where we recall $\phi_0$ is the bulk volume fraction of cations and anions each (with $0 < \phi_0 \le 1/2$). The specific case $\phi_0=1/2$ serves as a model for a solvent-free ionic liquid \cite{fedorov14}. The characteristic length $l=\sqrt{\nu/(4 \pi l_B)}$ in Eq.~\ref{fg54} reflects the volume $\nu$ per lattice site: we identify that volume with the ion volume. Eq.~\ref{ou41} implies for the Poisson-Fermi equation 
\begin{equation} \label{hh70}
f(\Psi)=\pm \sqrt{2 \ln \left[1+2 \phi_0 (\cosh \Psi -1)\right]},
\end{equation}
and thus $f^{-1}(s)=\mbox{arcosh}[1+(e^{s^2/2}-1)/(2 \phi_0)]$. With that we plot in Fig.~\ref{fig1} scaled curvature elastic constants for the Poisson-Fermi (solid lines) and the classical PB model (dashed lines) for different choices of $\phi_0$. 
The limit $|s| \ll 1$ (referred to as Debye-H\"uckel regime) yields $\kappa \nu/(k_BT l^3)=3 s^2/(16 \sqrt{2 \phi_0^3})$, $-\bar{\kappa}/\kappa=2/3$, and $c_0 l=2 \sqrt{2 \phi_0}/3$. In the opposite limit, $|s| \gg 1$, the diffuse part of the EDL becomes irrelevant, leaving layers of tightly condensed counterions that neutralize the surface charges. With $f(\Psi)=\sqrt{2 |\Psi|}$ we obtain from Eqs.~\ref{ft65} $\kappa \nu/(k_BT l^3)=2 |s|^5/15$, $-\bar{\kappa}/\kappa=1/4$, and $c_0 l=5/(8 |s|)$. Hence, accounting for the non-vanishing ion volume $\nu$ turns the saturation of $\kappa$ (and similarly for $\bar{\kappa}$), predicted in the PB limit, into growth $\sim |\sigma|^5$, irrespective of $\phi_0$. As a numerical illustration consider $\nu=1 \: \mbox{nm}^3$, $l_B=1 \: \mbox{nm}$, and $\sigma/e=1.7/\mbox{nm}^2$. This corresponds to $s=\sqrt{4 \pi l_B \nu}\: \sigma/e=6$. Because of $s \gg 1$, we find $\kappa/k_BT=(8 \pi l_B/15) \: \nu^3 (\sigma/e)^5=23$. Also, the non-vanishing ion volume tends to suppress instability with respect to spherical curvature, $c_1=c_2$. To this end, note that Eq.~\ref{ft58} implies the stability condition $-\bar{\kappa}/\kappa<2$. The PB limit predicts an instability for any choice of $\phi_0$, given $|s|$ is sufficiently large (see the dashed lines in Fig.~\ref{fig1}B). In contrast, the Poisson-Fermi model predicts an instability only for $\phi_0 \lesssim 0.002$, starting at about $s \approx 1.3$. 

\begin{figure}[ht!]
\begin{center}
\includegraphics[width=8.5cm]{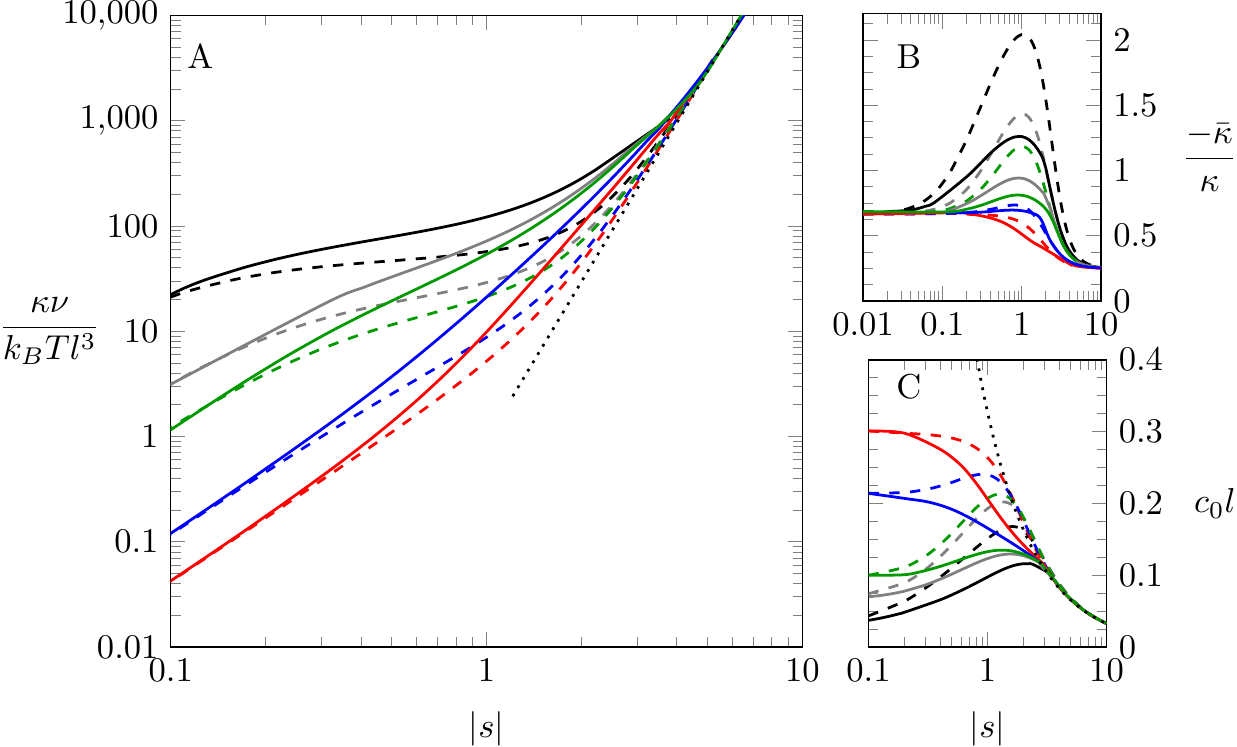}
\end{center}
\caption{\label{fig2} 
$\kappa/k_BT \times \nu/l^3$ (diagram A), $-\bar{\kappa}/\kappa$ (B), and $c_0 l$ (C), for $\phi_0=0.1, 0.05, 0.01, 0.005, 0.001$ (red, blue, green, grey, black) according to the Poisson-Carnahan-Starling model (solid lines) and the Poisson-Fermi model (dashed lines). The black dotted lines mark the large-$s$ limit. The Poisson-Fermi model is adjusted so that each spherical ion occupies a volume fraction $\pi/6$ of a lattice site.}
\end{figure}

Our method in Eq.~\ref{ft65} to calculate the curvature elastic constants is viable even when an analytic expression for $f(\Psi)$ is not available. For example, consider a class of mean-field models that assume the same particle size and shape for the mobile cations and anions, with an additional nonideality contribution added to the underlying equation of state.  The free energy of such a model can be expressed as the sum of the energy stored in the electric field and a mixing contribution corresponding to variations in the local volume fractions, $\phi_c$ and $\phi_a$, of the mobile cations and anions, respectively, 
\begin{eqnarray} \label{lo81}
\frac{F}{k_BT}&=&\frac{1}{\nu} \int \limits_V dv \Bigg[\frac{l^2}{2} {\left(\nabla \Psi\right)}^2+g_{id}(\phi_c)+g_{id}(\phi_a)\\
&+&g(\phi_c+\phi_a)-g(2 \phi_0)-(\phi_c+\phi_a-2 \phi_0) g'(2 \phi_0)\Bigg],\nonumber
\end{eqnarray}
where $g_{id}(\phi)=\phi \ln (\phi/\phi_0)-\phi+\phi_0$ is the ideal mixing free energy of the mobile ions and $g(\phi_c+\phi_a)$ is an additional nonideal contribution. 
The latter appears in the thermal equation of state of a homogeneous fluid with $N$ particles confined to a volume $V$ at pressure $P$ and temperature $T$ as $PV/(N k_BT)=1+g'(\phi)-g(\phi)/\phi$, where $g'(\phi)$ denotes 
the derivative with respect to the volume fraction $\phi=\nu N/V$. 
Variation of Eq.~\ref{lo81} yields the relations $\ln (\phi_c/\phi_0)=-\Psi-g'(\phi_c+\phi_a)+g'(2 \phi_0)$ and $\ln (\phi_a/\phi_0)=\Psi-g'(\phi_c+\phi_a)+g'(2 \phi_0)$ that define the equilibrium distributions $\phi_c=\phi_c(\Psi)$ and $\phi_a=\phi_a(\Psi)$. Generally, these are neither Boltzmann- nor Fermi-distributed; we can express them using the function $h(\phi)=\phi e^{g'(\phi)}$ and its inverse function $h^{-1}$ as
\begin{equation} \label{ji73}
\phi_{c/a}=\phi_0 e^{\mp \Psi} \frac{h^{-1}(h(2 \phi_0) \cosh \Psi)}{2 \phi_0 \cosh \Psi}.
\end{equation}
Using these in Poisson's equation $l^2 \nabla^2 \Psi=\phi_a-\phi_c$ yields the self-consistency relation $l^2 \nabla^2 \Psi=\tanh \Psi \times h^{-1}(h(2 \phi_0) \cosh \Psi)$. With Eq.~\ref{ou41} this gives rise to
\begin{equation} \label{jo87}
f(\Psi)=\pm \sqrt{2 \int \limits_0^\Psi d\bar{\Psi} \tanh \bar{\Psi} \times h^{-1}(h(2 \phi_0) \cosh \bar{\Psi} )}.
\end{equation}
When the function $h^{-1}$ is available in analytic form, $f(\Psi)$ may be obtained explicitly. An example is the Poisson-Fermi formalism discussed above: $g(\phi)=\phi+(1-\phi) \ln (1-\phi)$, implying $h(\phi)=\phi/(1-\phi)$ and $h^{-1}=\phi(h)=h/(1+h)$. Using these in Eq.~\ref{jo87}, we indeed recover Eq.~\ref{hh70}. Another example is the Carnahan-Starling equation of state, $PV/(N k_BT)=(1+\phi+\phi^2-\phi^3)/(1-\phi)^3$, and thus $g(\phi)=\phi^2 (4-3 \phi)/(1-\phi)^2$, as a model for an underlying hard-sphere fluid of mobile ions (all of equal size). Here, an analytic expression for $h^{-1}=\phi(h)$ is not available, but $h^{-1}$ can be computed numerically and then used to find $f(\Psi)$ according to Eq.~\ref{jo87}. Fig.~\ref{fig2} shows a comparison of predictions from the Poisson-Carnahan-Starling (solid lines) and Poisson-Fermi models (dashed lines).
For a meaningful comparison we adjusted the Poisson-Fermi model such that each mobile ion is spherical and thus occupies a volume fraction $\alpha=\pi/6$ of a cubic lattice site; this replaces Eq.~\ref{hh70} by $f=\pm \sqrt{2 \alpha \ln[1+2 \phi_0 (\cosh \Psi-1)/\alpha]}$. The differences observed in Fig.~\ref{fig2} for intermediate $s$ result from the higher pressure predicted by the Carnahan-Starling equation of state as compared to a lattice-gas. For example, the former has a second virial coefficient $4 \pi/3$ times larger than the latter.

  While Eq.~\ref{jo87} is restricted to ions of identical size and shape, Maggs and Podgornik \cite{maggs16} have recently made the connection of our function $f(\Psi)$ to the underlying electrolyte's equation of state for the general case of arbitrary ion sizes. Their analysis leads to $f(\Psi)=\sqrt{2 \nu \: \triangle P(\Psi)/k_BT}$, where $\triangle P$ is the excess osmotic pressure of the ions. For example, classical PB theory implies $\triangle P=2 \phi_0 (\cosh \Psi-1) k_BT/\nu$, and the symmetric lattice gas gives rise to $\triangle P=\ln [1+2 \phi_0 (\cosh \Psi-1)] k_BT/\nu$. Ref.~\citenum{maggs16} also discusses the extraction of the pressure for two size-asymmetric models, the Flory-Huggins and the Boublik-Mansoori-Carnahan-Starling-Leland equations of state. Eqs.~\ref{ft65} of our present work thus allow for the extraction of the curvature elastic constants according to these models.

Our final example is an extension of the Poisson-Fermi model, proposed by Han {\em et al} \cite{han14}, to anions and cations with mismatching volumes $\nu_c=\xi\nu$ and $\nu_a=\nu$, respectively, leading to the relation 
\begin{equation}\label{us29}
e^{\frac{\xi}{2} f(\Psi)^2}=\xi \phi_0 e^{-\Psi}+
\frac{\left[1-\phi_0 \left(1+\xi-e^{\Psi}\right)\right]^{\xi}}{(1-\xi \phi_0)^{\xi-1}}
\end{equation}
for the function $f(\Psi)$ defined in Eq.~\ref{hu35}. Here, the limiting behavior in the Debye-H\"uckel regime, $|s| \ll 1$, is $\kappa \nu/(k_BT l^3)=3 s^2/(16 \sqrt{2 \phi_{eff}^3})$, $-\bar{\kappa}/\kappa=2/3$, $c_0 l=2 \sqrt{2 \phi_{eff}}/3$, with the effective volume fraction $\phi_{eff}=\phi_0 [1-\phi_0 (1+\xi)/2]/(1-\xi \phi_0)$. The different ion sizes introduce asymmetry for positive and negative $\sigma$: for $-s \gg 1$ we obtain $\kappa \nu/(k_BT l^3)=2 |s|^5 \xi^3/15$, and $c_0 l=5/(8 \xi |s|)$, and for $s \gg 1$ we obtain $\kappa \nu/(k_BT l^3)=2 s^5/15$, and $c_0 l=5/(8 s)$. In both cases, $-\bar{\kappa}/\kappa=1/4$. 
\begin{figure}[ht!]
\vspace*{0.02cm}
\begin{center}
\includegraphics[width=8.0cm]{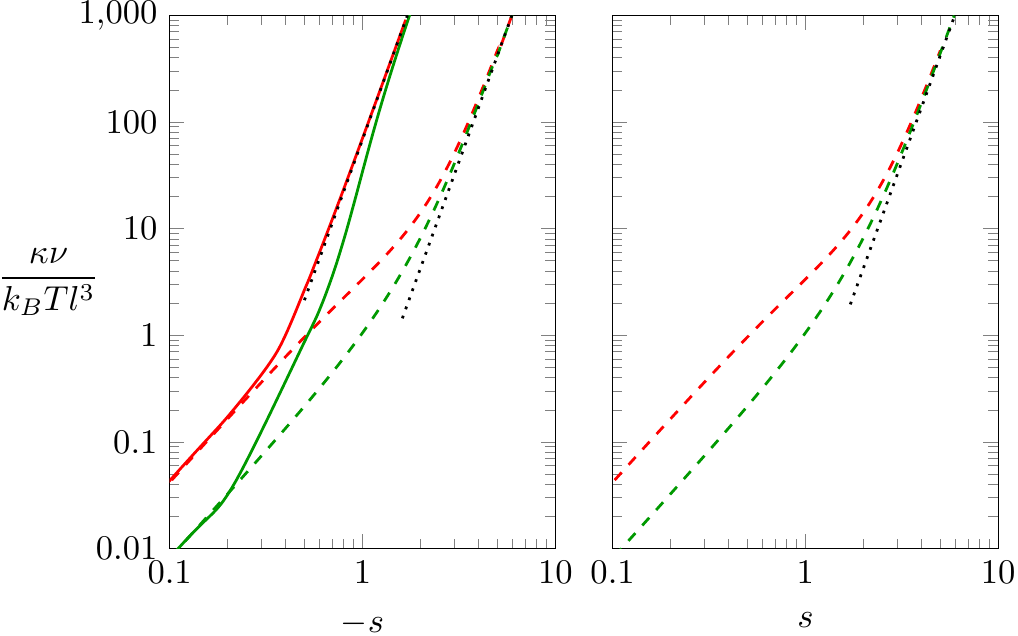}
\end{center}
\caption{\label{fig3} $\kappa/k_BT \times \nu/l^3$ for $\phi_0=0.1$ (colored red) and $\phi_0=0.3$ (green), computed for $\xi=1$ (dashed lines) and $\xi=2^3=8$ (solid lines). The dotted black lines mark the large-$s$ limit. Asymmetry for negative (left diagram) and positive (right diagram) $s$ emerges from the mismatching ion volumes $\nu_c=\xi\nu$ and $\nu_a=\nu$ (solid lines). Calculations are based on Eq.~\ref{us29}.
}
\end{figure}
Fig.~\ref{fig3} shows $\kappa \nu/(k_BT l^3)$ with its asymmetry for $s<0$ (left diagram) and $s>0$ (right diagram) for $\xi=2^3=8$ (solid lines). For comparison, we also display the case $\xi=1$ (dashed lines), for which $\kappa(s)=\kappa(-s)$.

In summary, we have introduced a general method to compute the curvature elastic moduli for a class of EDL models described by Eq.~\ref{hu35} and exemplified our approach based on both a lattice-gas (with and without mismatching ion sizes) and the Carnahan-Starling equation of state. Given the recently stated general relationship between Eq.~\ref{hu35} and the underlying equation of state of the bulk electrolyte \cite{maggs16}, it is now possible to include curvature effects into the calculation of EDL free energies. Our method leading to Eq.~\ref{ft65} can also be applied to electrodes with fixed surface potential, extended to arbitrary position of the neutral surface \cite{winterhalter92}, used to calculate the curvature dependence of the differential capacitance, and generalized to incorporate non-electrostatic, hydration-mediated ion-ion and ion-surface interactions.

G.V.~Bossa acknowledges a doctoral scholarship from CAPES Foundation (Grant No. 9466/13-4). 


\end{document}